\documentclass[aps,manuscript,showpacs,amsmath,amssymb,superscriptaddress]{revtex4-2}

\usepackage{tabularx}
\usepackage{bm}
\usepackage{graphicx}
\usepackage[scr=esstix]{mathalfa}

\usepackage{hyperref}
\hypersetup{colorlinks=true,urlcolor= blue,citecolor=blue,linkcolor= blue,bookmarksopen=false}

\usepackage{color}

\usepackage{amsmath,mathtools}
\usepackage{multirow}
\usepackage{dcolumn}
\usepackage{amssymb,amscd,xypic,bm,wasysym}
\usepackage{float}
\usepackage{cleveref}
\usepackage[caption=false,position=top,captionskip=0pt,farskip=0pt]{subfig}
\usepackage{soul}

\begin{document}
	
\title{Bounds and anomalies of inhomogeneous anomalous Hall effects}
	
\author{Christopher Ard}
\affiliation{Department of Physics, Colorado State University, Fort Collins, CO 80523, USA}
\author{Evan Camrud}
\affiliation{Department of Mathematics and Statistics, Middlebury College, Middlebury, VT 05753, USA}
\author{Olivier Pinaud}
\affiliation{Department of Mathematics, Colorado State University, Fort Collins, CO 80523, USA}
\author{Hua Chen}
\affiliation{Department of Physics, Colorado State University, Fort Collins, CO 80523, USA}
\affiliation{School of Materials Science and Engineering, Colorado State University, Fort Collins, CO 80523, USA}

\begin{abstract}
It is well recognized that interpreting transport experiment results can be challenging when the samples being measured are spatially nonuniform. However, quantitative understanding on the differences between measured and actual transport coefficients, especially the Hall effects, in inhomogeneous systems is lacking. In this work we use homogenization theory to find exact bounds of the measured or homogenized anomalous Hall conductivity (AHC) in inhomogeneous conductors under minimal assumptions. In particular, we prove that the homogenized AHC cannot exceed the bounds of the local AHC. However, in common experimental setups, anomalies that \emph{appear} to violate the above bounds can occur, with a popular example being the ``humps'' or ``dips'' of the Hall hysteresis curves usually ascribed to the topological Hall effect (THE). We give two examples showing how such apparent anomalies could be caused by different types of inhomogeneities and discuss their relevance in experiments.
\end{abstract}

\maketitle

\textit{Introduction---}Transport experiments are usually performed using macroscopic or mesoscopic samples that have unavoidable spatial inhomogeneities. Inhomogeneities on the scale much larger than the mean free path of the microscopic quasiparticles can be described by classical transport equations with spatially varying coefficients. The effective transport coefficients measured experimentally by probes located at the sample boundary or in the interior but separated by macroscopic distances \cite{pauw1958method} are, however, not trivially determined by the local ones through a simple spatial average \cite{pippard_book}. In certain cases the mesoscopic spatial inhomogeneity may lead to dramatic consequences not expected from the local physics. A prominent example is the classical inhomogeneity-induced non-saturating magnetoresistance \cite{Parish2003,Parish2005} proposed to explain the experimental observation in silver chalcogenides \cite{Xu1997}. 

Another anomaly in transport experiments that has attracted significant interest recently is the hump or dip features in Hall resistivity versus magnetic field curves \cite{kanazawa_large_2011,li_2013,Matsunoe1600304,soumyanarayanan_tunable_2017,maccariello_electrical_2018-1,zeissler_discrete_2018,raju_evolution_2019,ohuchi_topological_2015,shao_topological_2019,ahmed_spin-hall_2019,pang_spin-glass-like_2017-1,ohuchi_electric-field_2018,sohn_stable_2021,wang_ferroelectrically_2018,qin_emergence_2019,meng_observation_2019,gu_interfacial_2019,wang_spin_2019,sohn_hump-like_2020,zhang_robust_2020,wang_topological_2020-2,kim_2021,huang_2020,li_reversible_2020,vistoli_giant_2019-1,raju_colossal_2021,sapozhnikov_direct_2021,jiang_concurrence_2020,ren_topological_2024,kimbell_challenges_2022-1,He_2023,zhao_2022,chen_2020,algarni_2022,yan_2023,yu_2019,kumar_2020,sen_2019,kan_2020,Wysocki_2020,jia_unconventional_2021,liu_dimensional_2017,cheng_evidence_2019,vir_anisotropic_2019,pang_spin-glass-like_2017-1,spencer_helical_2018-1,wang_chiral-bubble-induced_2021,roy_intermixing_2015,tian_2021,gallagher_robust_2017,yokouchi_stability_2014,malsch_correlating_2020}, usually observed in magnetic conductors lacking spatial inversion symmetry either in the bulk or due to the presence of interfaces. A physically appealing picture of such anomalies is that finite magnetic fields stabilize an intermediate skyrmion phase in inversion-symmetry-breaking systems \cite{muhlbauer_2009}. It has been well established that conduction electrons coupled to skyrmion magnetic textures experience an emergent magnetic field and consequently have an extra Hall effect contribution, now known as the topological Hall effect (THE) \cite{ye_berry_1999,bruno_topological_2004,taguchi_2001,onoda_2004,Nagaosa2013}. The THE in the intermediate skyrmion phase therefore serves as a straightforward explanation for the anomalies \cite{Nagaosa2013}. However, such an interpretation, particularly when there is no complementing evidence from other experimental probes, has been seriously challenged in recent years \cite{kan_2018, gerber_2018, Groenendijk_2020, fijalkowski_coexistence_2020, kimbell_2020, kim_inhomogeneous_2020, yang2020origin,kan_2020,Wysocki_2020,chen_2020,algarni_2022,yan_2023, miao_strain_2020-1}. Among the alternative mechanisms, a widely accepted one is that superposition of anomalous Hall effect (AHE) \cite{Hall_1881,Karplus_1954,Smit_1958,Berger_1970,ye_berry_1999,Jungwirth_AHE_DMS_2002,Shindou_AHE_fcc_2001,nagaosa_2010,Tomizawa_2009,Chen_2014,Kubler_2014,Nakatsuji_2015} hysteresis loops coming from subsystems with distinct magnetic (e.g. coercive fields) or transport (e.g. signs of the AHE) properties in the measured samples can lead to similar hump or dip features \cite{kan_2018,kimbell_2020,kim_inhomogeneous_2020, wang_interface-induced_2021, kimbell_challenges_2022-1, gerber_2018, Groenendijk_2020, fijalkowski_coexistence_2020, wang_controllable_2020, liu_interface-induced_2024}. Effectively such a superposition is equivalent to performing a simple spatial average of the AHE. However, since the anomalies of the Hall signal occur near magnetization reversal which is a first-order phase transition, inhomogeneities due to formation of magnetic domains are prevalent. It is not clear if the measured anomalous Hall signals in an inhomogeneous magnetic conductor are always equivalent to such a spatial average. Moreover, the assumption of subsystems with individually well-defined coercive fields may be overly stringent \cite{sohn_hump-like_2020}. 

In the above context, it is desirable if one can constrain the measured Hall coefficients, especially the anomalous Hall conductivities (AHC), in inhomogeneous systems without assuming particular mechanisms, so that influences due to inhomogeneities only may be isolated. Theoretical bounds of AHC of nonuniform systems can also be a useful diagnostic tool for uncovering details of inhomogeneities by standard transport experiments. Exact results of such type are, however, relatively scarce in the physics literature (see e.g. \cite{Isichenko_1992,milton_2002} and references therein), possibly due to the general complexity of inhomogeneous transport problems and to the lack of experimental techniques for accurately characterizing inhomogeneity.

In this Article, we use homogenization theory, a topical field in partial differential equations (PDE), to study the bounds and anomalies of the measured AHE in inhomogeneous quasi-two-dimensional systems. This is enabled by assuming the system has well-defined local conductivities determined by microscopic electronic structure details, so that classical transport equations are applicable and the inhomogeneities are encoded in the spatially varying transport coefficients. We have first given exact bounds of the homogenized AHC in systems with both inhomogeneous longitudinal and Hall conductivities. The insights from this exact result further guide us to propose two inhomogeneity-related mechanisms that can lead to the hump/dip features in Hall hysteresis loops.

\textit{Bounds of the homogenized AHC---}We consider the following set of equations governing classical electric transport in 2D:
\begin{eqnarray}\label{eq:transportE}
\nabla \cdot \left(\pmb \sigma \cdot \mathbf E \right) = 0,\; \nabla \times \mathbf E = 0,
\end{eqnarray}
where 
\begin{eqnarray}
\pmb \sigma = \begin{pmatrix}
\sigma_{xx} & \sigma_{xy} \\
\sigma_{yx} & \sigma_{yy}
\end{pmatrix}
\end{eqnarray}
is the 2D conductivity tensor varying with position $\mathbf r = (x,y)$. The Hall conductivity $\sigma_{\rm h}$ is the antisymmetric part of $\bm \sigma$, i.e., $\sigma_{\rm h} \equiv (\sigma_{yx} - \sigma_{xy})/2$, and is due to either external magnetic fields (ordinary Hall effect) or spontaneous time-reversal-symmetry breaking (anomalous Hall effect). In this work we consider the latter only. $\sigma_{\rm h}$ is therefore odd under the reversal of the magnetic order parameter that breaks time-reversal symmetry, and fluctuates between positive and negative values in the presence of magnetic domains. 

The boundary condition is chosen so that the spatial average of $\mathbf E$ is equal to that of the externally applied electric field $\mathbf E_0$, i.e., $\langle \mathbf E \rangle \equiv \frac{1}{A}\int d^2\mathbf r \mathbf E =  \mathbf E_0$, $A$ being the sample area. The homogenized conductivity $\bar{\pmb \sigma}$ focused on in this work is defined through the spatial average of the electric current density $\mathbf j$ satisfying
\begin{eqnarray}\label{eq:sigmaeff}
\langle \mathbf j\rangle = \langle \pmb \sigma \cdot \mathbf E \rangle \equiv \bar{\pmb \sigma}\cdot \langle \mathbf E \rangle =  \bar{\pmb \sigma}\cdot \mathbf E_0.
\end{eqnarray}

To get some feel of the problem we start from the simplest case when $\sigma_{xx} = \sigma_{yy} \equiv \sigma_0$, $\sigma_{yx} =-\sigma_{xy} = \sigma_{\rm h} (\mathbf r)$, and $|\sigma_{\rm h} (\mathbf r)| = \sigma_{\rm h}$. Namely, only the AHC fluctuates between $\pm \sigma_{\rm h}$ spatially according to the magnetic domain profile. Bounds of $\bar{\sigma}_{\rm h}$ in this case can be obtained by first transforming $\pmb \sigma$ to an isotropic tensor $\pmb \sigma'$ using the duality transformation \cite{keller_1964, dykhne_1971_1, dykhne_1971_2, mendelson_1975, milton_1988}, and then applying the Hashin-Shtrikman bounds for 2-phase isotropic composites \cite{hashin_1962, milton_2002, supp}
\begin{eqnarray}\label{eq:bounds1}
\frac{\bar{\sigma}_{\rm h}}{\sigma_{\rm h}} \in \left[ -1 + \frac{2p \sigma_0^2 }{\sigma_0^2 + (1-p)^2 \sigma_{\rm h}^2}, \;1 - \frac{2(1-p) \sigma_0^2}{\sigma_0^2 + p^2 \sigma_{\rm h}^2}  \right]
\end{eqnarray}
where $p$ is the area percentage of domains with $\sigma_{\rm h}(\mathbf r) = \sigma_{\rm h}$. The two bounds are realized by domain configurations of packed coated circular cells with the same core/shell ratio \cite{milton_2002}. Moreover, one can see that the absolute value of $\bar{\sigma}_{\rm h}$ can never exceed $\sigma_{\rm h}$. The two bounds reduce to the trivial spatial-average value $\langle \sigma_{\rm h} (\mathbf r) \rangle = (2p-1)\sigma_{\rm h}$ when $\sigma_{\rm h}/\sigma_0 \rightarrow 0$.

Can $\bar{\sigma}_{\rm h}$ exceed the bounds of local $\sigma_{\rm h}$ when the longitudinal conductivity is inhomogeneous and in particular has a nonzero correlation with $\sigma_{\rm h}$? A simpler version of the problem can be obtained by replacing $\sigma_0$ by a constant $\sigma_1$ in domains with $\sigma_{\rm h}(\mathbf r) = \sigma_{\rm h}$, and that by another value $\sigma_2$ in domains with $\sigma_{\rm h}(\mathbf r) = -\sigma_{\rm h}$. Such a situation could happen during a magnetization reversal, when individual domains first nucleate at the regions with lower conductivities or more defects (see below). Note the values of $\sigma_{\rm h}(\mathbf r)$ can be either positive- or negative-correlated with the longitudinal conductivity, depending on the material and on which half of the hysteresis loop is under consideration. The bounds of $\bar{\sigma}_{\rm h}$ in this case can be obtained using the same approach as 
\begin{eqnarray}\label{eq:bounds2}
\frac{\bar{\sigma}_{\rm h}}{\sigma_{\rm h}} \in& \Bigg[& -1 + \frac{8 p \sigma_2^2}{[(1-p)\sigma_1 + (1+p) \sigma_2]^2 + 4(1-p)^2 \sigma_{\rm h}^2},\nonumber \\
&& 1 - \frac{8 (1-p) \sigma_1^2}{[(2-p)\sigma_1 + p \sigma_2]^2 + 4p^2 \sigma_{\rm h}^2} \Bigg].
\end{eqnarray}
Unlike Eq.~\eqref{eq:bounds1}, where the spatial average $\langle \sigma_{\rm h} \rangle$ is always in-between the two bounds, $\langle \sigma_{\rm h} \rangle$ can now fall outside the range in Eq.~\eqref{eq:bounds2}, demonstrating the nontrivial consequences of correlation between longitudinal and Hall conductivities. However, the bounds in Eq.~\eqref{eq:bounds2} still cannot go beyond the maximum or minimum of $\sigma_{\rm h}$.

The above observations, plus numerical results not shown here, motivated us to speculate that the homogenized AHC is always bounded by the maximum and minimum of the local AHC. This indeed turns out to be the case as we prove below with minimal assumptions. To be precise, we consider $\pmb \sigma(\mathbf r) = \pmb \sigma_0(\mathbf r) + \sigma_{\rm h}(\mathbf r) {\bf R}_\perp$, where $\pmb \sigma_0(\mathbf r)$ is a positive definite diagonal matrix with entries $\sigma_{xx}$ and $\sigma_{yy}$, and ${\bf R}_\perp$ is the $\pi/2$ rotation matrix in 2D. $\pmb \sigma_0$ and  $\sigma_{\rm h}$ can be correlated or not, and do not have to vary smoothly with $\mathbf r$. The theorem below, whose proof is given in the supplemental material \cite{supp}, shows that in this case $\bar{\sigma}_{\rm h}$ is still bounded by the global minimum and maximum of $\sigma_{\rm h}(\mathbf r)$:

\textbf{Theorem 1}. Suppose that $\sigma_{xx}$, $\sigma_{yy}$ and $\sigma_{\rm h}$ are stationary ergodic random fields (see e.g. \cite{berlyand} for definitions) and that $\sigma_{xx}$, $\sigma_{yy}$ are both bounded above and below by positive constants. Suppose additionally that
\begin{eqnarray}
-\sigma_M \leq \sigma_{\rm h}(\mathbf r) \leq \sigma_M,
\end{eqnarray}
for some positive constant $\sigma_M$. 
Then, we have the inequality
\begin{eqnarray}
- \sigma_M \leq \bar \sigma_{\rm h} \leq \sigma_M.
\end{eqnarray}

Theorem 1 holds under fairly general conditions: the homogenization regime arises when (i) the length at which the random fluctuations in the conductivity become statistically independent is small compared to the sample size and (ii) these fluctuations have the same statistical distribution across the sample. An additional technical requirement is for the conductivity to be bounded. Similarly to the law of large numbers, the homogenization regime may not hold if the fluctuations do not decorrelate sufficiently fast or if they have significantly different statistics across the sample.

Theorem 1 is independent of physical mechanisms contributing to the local $\bm \sigma$ and can be viewed as a no-go theorem for AHC anomalies induced by any types of inhomogeneities that can be captured by the spatially varying $\bm \sigma$. Practically, it suggests that in order for the measured AHC in the Hall hysteresis loop to exceed the values at saturation, there are only two possibilities: (1) The AHC at saturating magnetic fields does not correspond to the bounds of the local AHC in the middle of the hysteresis; (2) The measured quantity is not the homogenized AHC. Note that (1) also incorporates the skyrmion scenario. Motivated by this classification, in the following two sections we provide examples for both possibilities but not due to skyrmions.

\textit{Anomalies due to nonuniform saturated AHC---}In this section we discuss a possible mechanism for the anomalies when the spatial dependence of $\sigma_{\rm h}(\mathbf r)$ is not only due to the magnetic domains. To this end we first give a recipe for modeling magnetic domain evolution in field sweeps. As the perpendicular magnetic field sweeps, e.g., from negative to positive values, one expects the magnetic configuration to continuously evolve from a uniform negative magnetization, to spatially separated positive/negative-magnetization domains in a relatively narrow field range centered at the coercive field $H_c$, and finally to a uniform positive magnetization. This process can be qualitatively captured by the following function describing the spatial profile of normalized perpendicular magnetization
\begin{eqnarray}\label{eq:mprofile}
m(\mathbf r) = {\rm sgn}\left[\frac{H - H_c}{\epsilon_H} + M(\mathbf r) \right]
\end{eqnarray}
where ${\rm sgn}(x) = 1$ $(-1)$ when $x>0$ $(x<0)$, $H$ is the perpendicular magnetic field, $H_c$ is the coercive field, $M(\mathbf r)$ is a smooth random field ranging from $M_{\rm min}<0$ to $M_{\rm max}>0$, and $\epsilon_H$ controls the abruptness of magnetic switching. For analytical convenience, below we replace ${\rm sgn}(x)$ by ${\rm erf}(x)$ where ${\rm erf}(x) = \frac{2}{\sqrt{\pi}}\int_{0}^{x} e^{-t^2} dt$ is the error function. Note that Eq.~\eqref{eq:mprofile} gives infinitely sharp domain walls, while using the error function makes the domain wall width be determined by the $\mathbf r$ dependence of $M$.  

If the inhomogeneity of $\sigma_{\rm h}(\mathbf r)$ is only caused by magnetic domains, one can express it simply as $\sigma_{\rm h}(\mathbf r) = \sigma_{\rm h} m(\mathbf r)$, where $\sigma_{\rm h}$ is the absolute value of the AHC in magnetically saturated states. However, in the presence of inhomogeneous material composition in real samples, even if the local magnetization in different spatial regions is the same, the local AHC does not have to have the same value or sign. This is particularly the case for nearly compensated ferrimagnets whose AHC changes sign across the compensation point \cite{buschow_1980,choe_1987,hansen_1989,finley_2016,Xiao2021}. To account for such an inhomogeneity, we generalize $\sigma_{\rm h}(\mathbf r)$ to the following 
\begin{eqnarray}\label{eq:sigmahrand}
    \sigma_{\rm h} (\mathbf r) = \sigma_{\rm h}\left[1 + r_{\rm h} C(\mathbf r) \right] m(\mathbf r)
\end{eqnarray}
where $C(\mathbf r)$ ($C$ stands for composition) is a zero-mean random field whose typical size is controlled by the dimensionless number $r_{\rm h} > 0$.

While $C(\mathbf r)$ is in general distinct from $M(\mathbf r)$, it is reasonable to expect them to have certain correlation, as already mentioned in the last section. For example, the energy barrier for locally nucleating a magnetic domain in an inhomogeneous sample is ultimately determined by the local material composition. Therefore regions where the local magnetization flips first during field sweeps should also have consistently larger/smaller AHC compared to the other regions. Similar correlation can also exist between local AHC and other local magnetic properties such as coercivity and saturation magnetization. 

We now show that the correlation between $C(\mathbf r)$ and $M(\mathbf r)$ generally leads to hump/dip features in Hall hysteresis loops, which already manifests in the lowest order approximation of the homogenized conductivity $\bar{\mathbf \sigma}\approx \langle \mathbf \sigma \rangle$. Namely,
\begin{eqnarray}\label{eq:sigmaheff1}
    \bar{\sigma}_{\rm h} \approx \langle \sigma_{\rm h} (\mathbf r)\rangle = \sigma_{\rm h}\left\langle \left(1 + r_{\rm h} C \right) {\rm erf}\left( \frac{H-H_c}{\epsilon_H} + M \right)\right\rangle 
\end{eqnarray}
To proceed, we assume both $C$ and $M$ are Gaussian random fields with the following variance and covariance
\begin{eqnarray}
\langle C^2\rangle = 1,\, \langle M^2\rangle = 1,\,  \langle CM\rangle = \mathcal{V}_{CM}  
\end{eqnarray}
where the fields in the angle brackets have the same position variable $\mathbf r$ and the covariance $\mathcal{V}_{CM}$ can be either positive or negative depending on the nature of the correlation. Eq.~\eqref{eq:sigmaheff1} can then be formally calculated using Wick's theorem \cite{supp}. However, before using the resulting formula to numerically evaluate $\langle{\sigma}_{\rm h}\rangle$, we derive some useful qualitative results first.

The extrema of $\langle{\sigma}_{\rm h}\rangle$ versus $\delta h \equiv \frac{H-H_c}{\epsilon_H}$ can be found by
\begin{eqnarray}
    0 = \frac{\partial\langle \sigma_{\rm h}\rangle }{\partial (\delta h)} \approx \frac{2}{\sqrt{\pi}}\sigma_{\rm h}\langle (1 + r_{\rm h} C) (1-2\delta h M - M^2)\rangle 
\end{eqnarray}
which gives the critical values $\delta h_c = 0$. The resulting extremal value of $\langle \sigma_{\rm h}\rangle$ is 
\begin{eqnarray}
    \langle \sigma_{\rm h} \rangle_c \approx \frac{2}{\sqrt{\pi}}\sigma_{\rm h} \langle (1 + r_{\rm h} C) (\delta h_c + M)\rangle
   = \frac{2}{\sqrt{\pi}} \sigma_{\rm h} r_{\rm h} \mathcal{V}_{CM} 
\end{eqnarray}
whose absolute value can exceed $\sigma_{\rm h}$ if $r_{\rm h}|\mathcal{V}_{CM}| \gtrsim 0.89$. Since $|\mathcal{V}_{CM}|$ is expected to be less than 1, $r_{\rm h}$ should be on the order of 1 or larger. This means that the fluctuation of $\sigma_{\rm h}$ must be strong enough so that its sign can change spatially even in a uniform magnetic state.

\begin{figure}[ht]
	\centering
 	\subfloat[]{\includegraphics[width=0.35\textwidth]{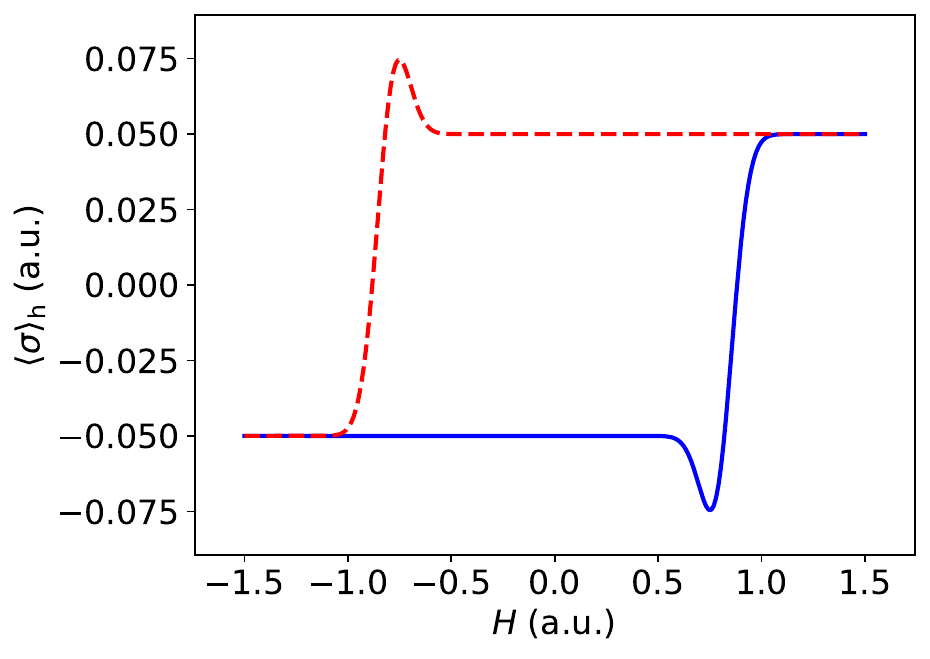}}
	\subfloat[]{\includegraphics[width=0.35\textwidth]{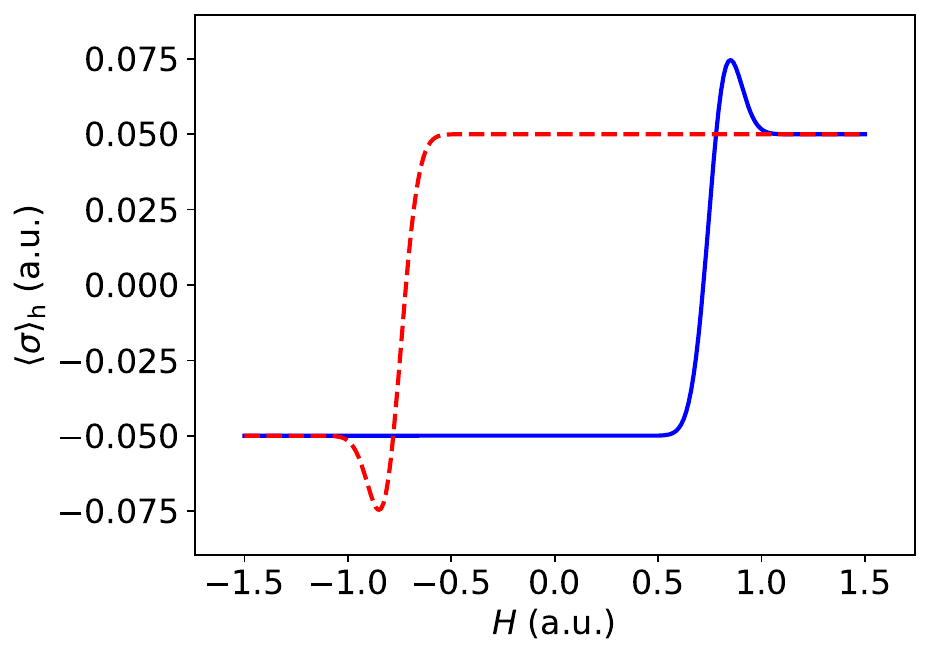}}
	\caption{Dip (a) and hump (b) features due to nonuniform saturated values of $\sigma_{\rm h}$ correlated with magnetic domains, plotted using Eq.~\eqref{eq:sigmaheff1}. The up-sweeping part of each loop is plotted using $\sigma_{\rm h} = 0.05$, $H_c = 0.8$, $\epsilon_H = 0.1$, $r_{\rm h} = 4.0$, $\langle C^2\rangle = 1$, $\langle M^2\rangle = 0.09$, $\mathcal{V}_{CM} = -0.3$ (a) and $\mathcal{V}_{CM} = 0.3$ (b), while the down-sweeping half is obtained by flipping the signs of $H_c$ and $\mathcal{V}_{CM}$.}
    \label{fig:hloop}
\end{figure}

The above qualitative picture is verified by Figure~\ref{fig:hloop}, which plots Eq.~\eqref{eq:sigmaheff1} versus $H$ as well as its time reversal, amounting to $(H_c, M)\rightarrow (-H_c, -M)$. The parameter values are chosen so that they are comparable to that in systems known to host multiple AHE channels \cite{wang_controllable_2020}. Explicitly calculating the Hall conductance using the finite element method as detailed in the next section also gives similar results \cite{supp}. Note that the half of hysteresis loop with positive $H_c$ corresponds to sweeping $H$ from negative to positive values. A hump in the upsweeping curve therefore occurs when $\mathcal{V}_{CM}>0$ [Fig.~\ref{fig:hloop} (b)]. More intuitively, when $H\approx H_c > 0$, regions with $M>0$ are positive domains nucleated within a negative magnetization background. The positive $\mathcal{V}_{CM}$ therefore means that regions with larger AHC are switched first. When a local coercive field can be approximately defined, it means regions with larger AHC have smaller coercive fields, as proposed in previous literature \cite{gerber_2018, kan_2018, Groenendijk_2020, fijalkowski_coexistence_2020}, but our theory applies to more general situations such as that in \cite{sohn_hump-like_2020, ren_topological_2024}. 

\textit{Anomalies due to domain wall resistance---}In this section we consider another possible cause of the anomalies, i.e., it is not the $\bar{\sigma}_{\rm h}$ that is actually measured. We also consider another complication that is not accounted for in Theorem 1, i.e., when $\bm \sigma(\mathbf r)$ has symmetric off-diagonal components. 

Experiments for Hall effects are usually performed using ``Hall bar'' devices, where the longitudinal currents flowing through the sample are externally controlled and are expected to be nearly uniform near the middle of the bar, and it is the voltage between transverse boundaries that is measured. Supposing one considers a system in the domain $x, y\in [0,1]$, the above measurement geometry can be described by the following boundary conditions:
\begin{equation}\label{eq:bcHallbar}
	\mathbf j\cdot \hat{n} =
	\begin{cases}
		-\sigma_0 E_0 , & x=0\\
		\sigma_0 E_0 , & x=1\\
		0, & y=0,1
	\end{cases}
\end{equation}
where $\hat{n}$ is the unit normal vector of the boundary pointing to the exterior of the solution domain. The combination of $\sigma_0 E_0$ for specifying the longitudinal current $I_x$ is introduced for later convenience. One can then obtain $R_{yx} = V_y/I_x$, where $V_y$ is the voltage difference between top and bottom edges averaged over $x$. The Hall resistance is $R_{yx}$ after antisymmetrization: $R_{\rm h} = (R_{yx} - {\mathcal{T}}[R_{yx}])/2$ where $\mathcal{T}$ means time reversal.

We solve the above problem for $R_{\rm h}$ by considering an anisotropic contribution to $\bm \sigma(\mathbf r)$ from the magnetic domain walls, which prevail during the magnetization reversal. It is well known \cite{levy_1997} that currents flowing parallel and perpendicular to a domain wall experience different resistivities. Near a domain wall oriented along the $y$-axis, we can define the domain wall contribution to $\bm \sigma$ as
\begin{equation}\label{eq:sdw}
	\bm \sigma_{\rm DW} = f(\mathbf r) \begin{pmatrix}
		\sigma_\perp & 0 \\
		0 & \sigma_\parallel
	\end{pmatrix}
\end{equation}
where $\sigma_\perp$ ($\sigma_\parallel$) is the local change to the $xx$ ($yy$) component of the conductivity tensor. Normally one expects $-\sigma_0 < \sigma_\perp < \sigma_\parallel < 0$. $f(\mathbf r)$ is a positive scalar function that smoothly increases from 0 to 1 as one moves from domain interior to the wall. Using notations in the last section, we can, for example, choose
\begin{eqnarray}
	f(\mathbf r) = 1 - {\rm erf}^2\left[ \frac{1}{\epsilon_W} \left( \frac{H - H_c}{\epsilon_H} + M \right)   \right]    
\end{eqnarray}
where the new parameter $\epsilon_W$ controls the typical width of $f$ variation relative to that of magnetization domain walls, since the spatial range in which $f(\mathbf r)$ becomes significant does not have to be the same as that of the magnetization variation near a domain wall. We can then rotate Eq.~\eqref{eq:sdw} to describe arbitrarily oriented domain walls as
\begin{equation}\label{eq:rsdw}
	\bm \sigma_{\rm DW}(\mathbf{r})
	=
	f(\mathbf{r})
	R [\theta(\mathbf r)]
	\begin{pmatrix}
		\sigma_\perp & 0 \\
		0 & \sigma_\parallel
	\end{pmatrix}
	R^{-1} [\theta(\mathbf r)]
\end{equation}
where $R [\theta(\mathbf r)]$ is the 2D rotation matrix with the rotation angle $\theta$ a function of $\mathbf r$. $\theta(\mathbf r)$ can be obtained from $f(\mathbf r)$ since the gradient of $f$ is perpendicular to the domain wall: $(\cos \theta, \sin\theta) = \frac{1}{|\nabla f|}(\partial_x f, \partial_y f)$. 

We next solve the transport equations with the boundary condition Eq.~\eqref{eq:bcHallbar}, $\bm \sigma_{\rm DW}$ in Eq.~\eqref{eq:rsdw}, and Eq.~\eqref{eq:sigmahrand} with $C=0$ for $\sigma_{\rm h}(\mathbf r)$. This is done numerically with FEniCS \cite{LoggMardalEtAl2012} and $M$ generated as a 2D Gaussian random field in the unit square. The time reversal of $R_{yx}$ for defining $R_{\rm h}$ corresponds to $(H, M, H_c)\rightarrow (-H, -M, -H_c)$ in our model.

\begin{figure}
    \center
     \subfloat[]{\includegraphics[width=0.3\textwidth]{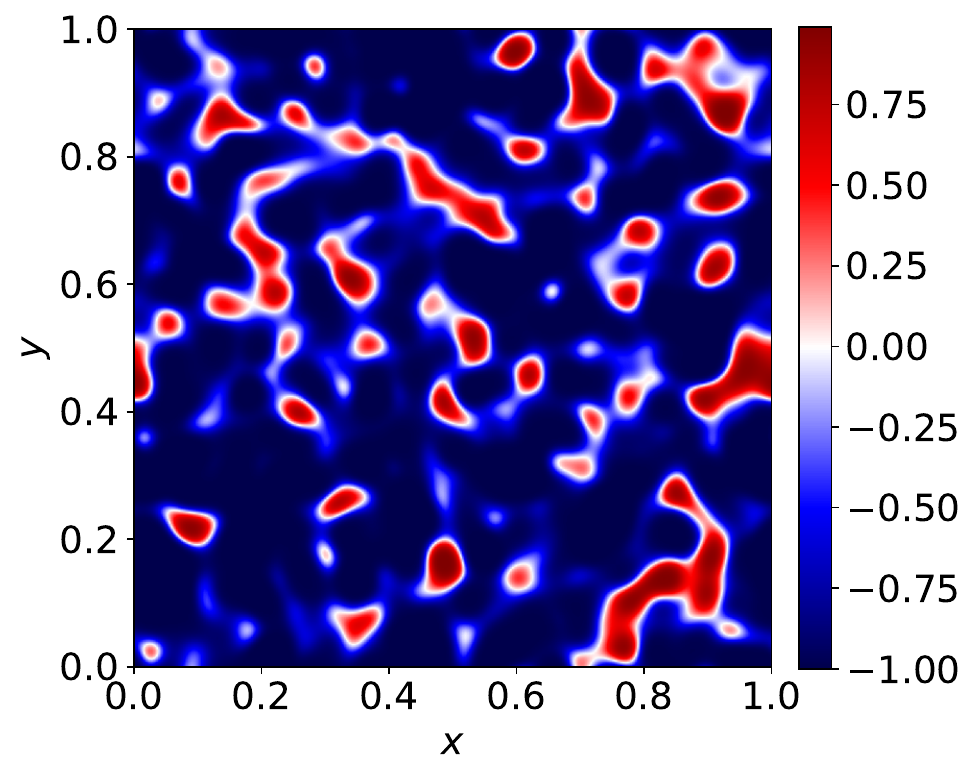}}\,
    \subfloat[]{\includegraphics[width=0.33\textwidth]{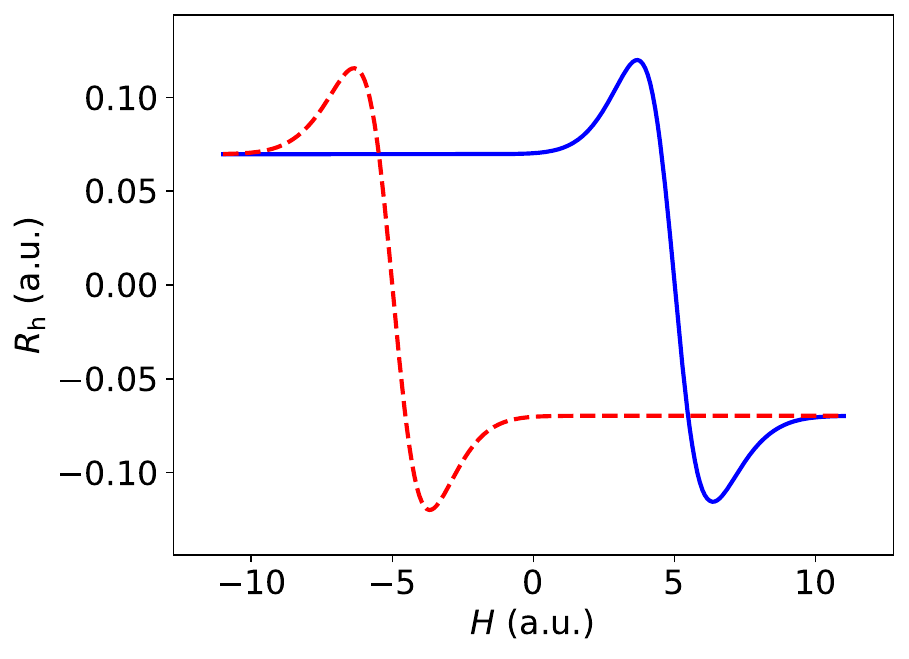}} \\
         \subfloat[]{\includegraphics[width=0.3\textwidth]{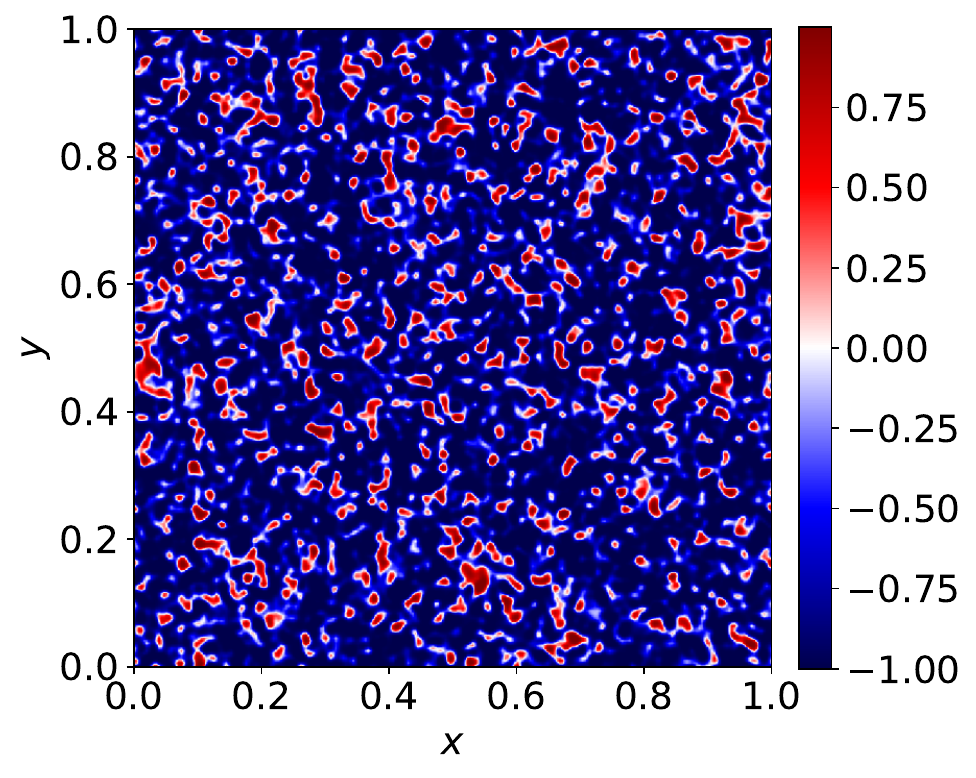}}\,
    \subfloat[]{\includegraphics[width=0.33\textwidth]{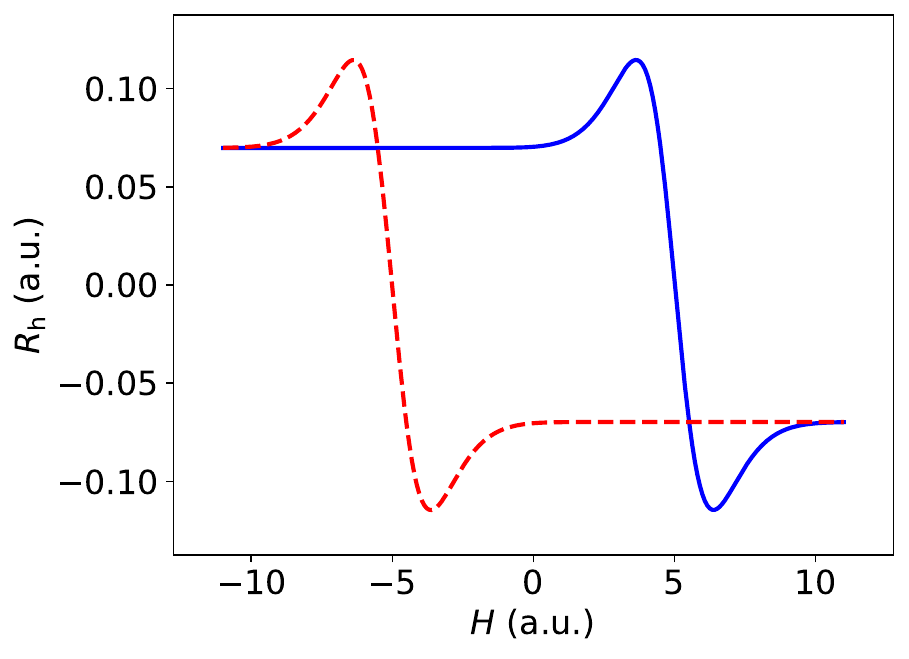}}
    \caption{(b, d) Hysteresis loops with humps/dips induced by the domain wall resistance mechanism, calculated using $\xi = 3\times 10^{-2}$ (a) and $\xi = 8\times 10^{-3}$ (c), respectively. (a, c) Domain configurations $m(\mathbf r)$ at $H = 3.9$ of up-sweeping curves in (a) and (c), respectively. The other parameter values are: $\sigma_{\rm h}=0.07, \sigma_\perp=-0.8, \sigma_\parallel = -0.4, H_c =5.0, \epsilon_{W} = 2.0$, $\sigma_0 = E_0 = \epsilon_{H} = 1.0$.}
    \label{fig:dw_hyst}
\end{figure}

Figure~\ref{fig:dw_hyst} (b) plots the hysteresis loop $R_{\rm h}(H)$ obtained from the above numerical scheme. The parameter values are chosen so that the maximal magnetoresistance is of similar order as that in \cite{jiang_concurrence_2020}. A distinct feature of the figures, different from Fig.~\ref{fig:hloop}, is that both a hump and a dip appear in a single half of the hysteresis. Figure~\ref{fig:dw_hyst} (a) shows the domain profile at the peak position of the up-sweeping curve, indicating that the correlation length $\xi$ of the random field $M$ is not very small compared to the solution domain. However, by further calculating configurations with $\xi \ll 1$, as exemplified by Fig.~\ref{fig:dw_hyst} (c), the double hump/dip feature still survives [Fig.~\ref{fig:dw_hyst} (d)], suggesting the domain wall mechanism is relevant even in the homogenization limit.

The origin of the double hump/dip feature in the homogenized Hall \emph{resistance} can be understood from a heuristic perturbation calculation as follows. Up to the lowest nonzero order in $\sigma_{\rm h} (\mathbf r)$, the solution of the transport equation is $\mathbf E\approx \mathbf E^{(0)} + \mathbf E^{(1)}$ where the superscript means the order in $\sigma_{\rm h}$. The spatial average of $\mathbf j(\mathbf r)$ is then
\begin{eqnarray}
    \langle \mathbf j \rangle \approx \left \langle (\sigma_0 \mathbf I + \sigma_{\rm h}\mathbf R_\perp + \bm \sigma_{\rm DW} ) \left( \mathbf E^{(0)} + \mathbf E^{(1)} \right)\right \rangle. 
\end{eqnarray}
For simplicity we approximate $\langle AB \rangle \approx \langle A\rangle \langle B \rangle$, and consider the homogenization limit when $\langle \sigma_{\rm DW}^{xx}\rangle = \langle \sigma_{\rm DW}^{yy}\rangle \equiv \sigma_{\rm DW}$ and $\langle \sigma_{\rm DW}^{xy}\rangle = \langle \sigma_{\rm DW}^{yx}\rangle = 0$. The boundary condition $\langle j_x \rangle = j_0$, $\langle j_y \rangle = 0$ then leads to
\begin{eqnarray}
     &&(\sigma_0 +  \sigma_{\rm DW}) \langle E_x^{(0)} +  E_x^{(1)}\rangle  - \langle \sigma_{\rm h}\rangle \langle E_y^{(1)}\rangle = j_0\\\nonumber 
    &&(\sigma_0 + \sigma_{\rm DW} )\langle E_y^{(1)}\rangle+ \langle \sigma_{\rm h}\rangle \langle E_x^{(0)} +  E_x^{(1)}\rangle= 0
\end{eqnarray}
from which we get the Hall resistivity
\begin{eqnarray}
    \bar{\rho}_{\rm h} \approx - \frac{\langle E_y^{(1)}\rangle}{j_0} = \frac{\langle \sigma_{\rm h}\rangle}{(\sigma_0 +\sigma_{\rm DW})^2 + \langle\sigma_{\rm h}\rangle^2}
\end{eqnarray}
consistent with that obtained by inverting the averaged conductivity tensor as expected. Compared to the saturated value $\rho_{\rm h} = \sigma_{\rm h}/(\sigma_0^2 + \sigma_{\rm h}^2)$, $\bar{\rho}_{\rm h}$ differs by a factor of $r_{\rm DW} \equiv \frac{\bar{\rho}_{\rm h}}{\rho_{\rm h}}\approx \frac{\langle \sigma_{\rm h} \rangle }{\sigma_{\rm h}}(1 - |  \sigma_{\rm DW} | /\sigma_0)^{-2}$, assuming the Hall conductivities are negligible compared to the longitudinal ones. $r_{\rm DW}$ is always positive when $\langle \sigma_{\rm h} \rangle$ has the same sign as the saturated value $\sigma_{\rm h}$. Therefore both a hump and a dip will appear simultaneously in either half of the hysteresis loop when $r_{\rm DW} > 1$, or equivalently $\frac{\langle \sigma_{xx}\rangle }{\sigma_0} < \sqrt{\frac{\langle \sigma_{\rm h}\rangle }{\sigma_{\rm h}}}$, at certain values of $H-H_c$. Since $\frac{\langle \sigma_{xx}\rangle }{\sigma_0}$ grows quadratically with $H-H_c$ near $H_c$, while $\frac{\langle \sigma_{\rm h}\rangle }{\sigma_{\rm h}}$ increases linearly with $H-H_c$, the inequality always holds when $\frac{\langle \sigma_{xx}\rangle (H_c) }{\sigma_0}\rightarrow 0$, i.e., when the longitudinal conductivity is strongly suppressed by the presence of domain walls. 

The coexistance of hump/dip in a single half of the hysteresis loop, together with abnormally large magnetoresistance near the magnetic switching, are strong indicators of our domain wall mechanism. Such a scenario has been briefly discussed in \cite{jiang_concurrence_2020}, where the difference between $\bar{\rho}_{\rm h}$ and $\bar{\sigma}_{\rm h}$ is also emphasized. We also note that when the inhomogeneous AHC size discussed in the last section coexists with the domain wall mechanism, the double hump/dip can have different magnitudes \cite{supp}. The double hump/dip features due to the domain wall mechanism can be very similar to that caused by the THE when the coercive field is negligible. In such cases the field dependence of the longitudinal resistance is essential for distinguishing the two. 

Finally, as shown in \cite{supp}, we have consistently observed that the Hall conductance, obtained by first inverting the resistance tensor and then antisymmetrizing the off-diagonal part, is always bounded by the saturated values in the homogenization limit \cite{jiang_concurrence_2020}. This strongly suggests that the conclusion of Theorem 1 still holds when the conductivity tensor has spatially dependent anisotropy.

\textit{Discussion---}Theorem 1 suggests that the anomalies of measured AHC can only appear if its value at saturating magnetic fields does not correspond to the extrema of the local AHC during magnetic reversal. Efforts in pinning down the causes of anomalies in a specific sample should therefore be focused on characterizing the structural inhomogeneity, as well as microscopic calculations for the size of the AHE in the presence of skyrmions or chiral domain walls. Since the theorem applies in the homogenization limit, it is also necessary to ensure that there is little sample dependence of the anomalies.

Our study is based on the classical transport equation with spatially varying coefficients. As a result, it also incorporates quantum effects as long as their contribution can be described as a local conductivity tensor, such as the THE as mentioned previously. Therefore our formalism can in principle be generalized to include other interesting quantum effects such as domain wall modes in Weyl semimetals, integer or fractional Chern insulators, etc., and to study their consequences in the presence of various mesoscopic inhomogeneities. Such a formalism does involve approximations, mainly the response to electromagnetic fields being local, and therefore holds when the mean free path is shorter than the length scale of the inhomogeneity. In the opposite limit, homogenization must be done first in the quantum transport theory. The resulting conductivity tensor can then be used in a similar fashion for the classical transport formalism to deal with inhomogeneities on a larger length scale. Additionally, since the same transport equation applies to a wide variety of steady-state transport phenomena, our results can be straightforwardly applied to them as well. One example is the thermal transport in chiral superconductors or quantum spin systems where a zero-field, i.e. anomalous, thermal Hall effect exists because of broken time-reversal symmetry.

Our study also calls for the need of developing transport techniques that can characterize the extent of inhomogeneity and allow for comparison with a full solution of the transport problem in a given system. This is in the same line as recent experiments in disentangling domain wall vs. bulk contributions to the Hall conductivity in quantum anomalous Hall systems \cite{rosen_2022,Ferguson2023}. 

\begin{acknowledgements} 
The authors thank Mingzhong Wu for helpful discussions. CA was supported by NSF CAREER grant DMS-1452349 and by NSF CAREER grant DMR-1945023. OP was supported by NSF CAREER grant DMS-1452349 and NSF grant DMS-2006416. HC was supported by NSF CAREER grant DMR-1945023. HC also thanks Aspen Center for Physics, which is supported by National Science Foundation grant PHY-2210452, where part of the work was performed.
\end{acknowledgements}

\bibliography{ahc_ref}
\end{document}